\documentclass[12pt]{article}

\newcommand{\lcc}{{\widehat\nabla}} 
\newcommand{\Klcc}{{\widehat K}} 
\newcommand{\Rlcc}{{\widehat R}} 
\newcommand{\bnw}{{\nabla}} 
\newcommand{\Rbnw}{{R}} 
\newcommand{\clcc}{{\overset{\mr c}{\nabla}}} 
\newcommand{\slcc}{\nabla^{\mr s}} 
\newcommand{\lvf}{\lambda} 

\usepackage{amsmath}
\usepackage{amssymb}
\usepackage{amsthm}
\usepackage{enumerate}
\usepackage{bbm}
\usepackage{bm}
\usepackage{mathtools}
\usepackage{tikz-cd}
\usepackage{hyperref}
\usepackage{xcolor}

\theoremstyle{theorem}
\newtheorem*{Theorem}{Theorem}
\newtheorem*{Proposition}{Proposition}
\newtheorem{Lemma}{Lemma}

\theoremstyle{remark}
\newtheorem*{Remark}{Remark}

\newcommand{\btm}{\begin{Theorem}}
\newcommand{\etm}{\end{Theorem}}

\newcommand{\ben}{\begin{enumerate}}
\newcommand{\een}{\end{enumerate}}

\newcommand{\bit}{\begin{itemize}}
\newcommand{\eit}{\end{itemize}}

\newcommand{\bca}{\begin{cases}}
\newcommand{\eca}{\end{cases}}

\newcommand{\bre}{\begin{Remark}\rm}
\newcommand{\ere}{\end{Remark}}

\newcommand*{\bbm}{\begin{Remark}}
\newcommand*{\ebm}{\end{Remark}}

\newcommand{\ble}{\begin{Lemma}}
\newcommand{\ele}{\end{Lemma}}

\newcommand*{\bsz}{\begin{Proposition}}
\newcommand*{\esz}{\end{Proposition}}

\newcommand{\beq}{\begin{equation}}
\newcommand{\eeq}{\end{equation}}

\newcommand{\bbma}{\begin{bmatrix}}
\newcommand{\ebma}{\end{bmatrix}}

\newcommand*{\bbs}{\begin{Example}}
\newcommand*{\ebs}{\end{Example}}

\newcommand*{\bfg}{\begin{Corollary}}
\newcommand*{\efg}{\end{Corollary}}

\newcommand*{\bdf}{\begin{Definition}}
\newcommand*{\edf}{\end{Definition}}

\newcommand*{\bbw}{\begin{proof}}
\newcommand*{\ebw}{\end{proof}}

\newcommand*{\bpf}{\begin{proof}}
\newcommand*{\epf}{\end{proof}}

\newtagform{simple}{}{}
\newtagform{standard}{(}{)}

\allowdisplaybreaks

\newcommand{\CC}{{\mathbb{C}}}

\newcommand{\RR}{{\mathbb{R}}}

\newcommand{\sitem}{\rm\item\it}

\newcommand{\al}[1]{\begin{align} #1 \end{align}}
\newcommand{\ala}[1]{\begin{align*} #1 \end{align*}}

\newcommand{\mr}[1]{\mathrm{#1}}

\newcommand{\comment}[1]{}
\newcommand{\verweis}[1]{}
\newcommand{\todo}[1]{}

\newcommand{\ctg}{\mr T^\ast}
\newcommand{\tg}{\mr T}

\newcommand{\rref}[1]{{\rm \ref{#1}}}

\newcommand{\abs}{\hspace*{2.5mm}}
\newcommand*{\qeb}{\nopagebreak\hspace*{0.1em}\hspace*{\fill}{\mbox{\small$\blacklozenge$}}}

\newcommand{\g}{\mathfrak{g}}

\newcommand{\calC}{\mathcal{C}}
\newcommand{\calD}{\mathcal{D}}
\newcommand{\calA}{\mathcal{A}}

\newcommand{\CE}{\operatorname{CE}}

\begin{document}

\title{On a connection used in deformation quantization}

\author{G.\ Rudolph, M.\ Schmidt
\\[5pt]
Institute for Theoretical Physics, University of Leipzig, 
\\
P.O. Box 100 920, D-4109 Leipzig, Germany}

\maketitle

\begin{abstract}

\noindent
We show that the connection used by Bordemann, Neumaier and Waldmann \cite{BNW} to construct the Fedosov standard ordered star product on the cotangent bundle of a Riemannian manifold is obtained by symplectification of the complete lift of the corresponding Levi-Civit\'a connection, in the sense of Yano and Patterson \cite{YP1}. In terms of local coordinates, this was shown already earlier by Pleba\'nski, Przanowski and Turrubiates \cite{PPT}. In the final part, we comment on the usefulness of this connection in the context of homological reduction of deformation quantized models.

\end{abstract}

\section{Introduction}

In Fedosov deformation quantization \cite{Fedosov} of models living on the cotangent bundle $\ctg Q$ over a Riemannian manifold $Q$ one uses a lift of the Levi-Civita connection to $\ctg Q$ which is torsionless, symplectic and homogeneous to construct a so-called Fedosov derivation. The latter is crucial for the definition of the Fedosov star product defining the deformation quantization structure. 

In more detail, let $Q$ be a manifold endowed with a Riemannian metric $g$ and let $\lcc$ denote the corresponding Levi-Civit\'a connection. Denote by $\ctg Q$ the cotangent bundle of $Q$, by $\pi : \ctg Q \to Q$ the natural projection, by $\theta$ the tautological $1$-form on $\ctg Q$, and by $\omega :=\mr d \theta$ the canonical symplectic form. Recall that a torsion-free linear connection $\bnw$ on $\ctg Q$ is called

\ben

\item a lift of $\lcc$ if
$
\pi' \circ (\bnw_X Y)
=
(\lcc_{\hat X} {\hat Y}) \circ \pi
$
for all vector fields $X$, $Y$ on $\ctg Q$ and $\hat X$, $\hat Y$ on $Q$ satisfying $\pi' \circ X = \hat X \circ \pi$ and $\pi' \circ Y = \hat Y \circ \pi$,

\item symplectic if
$
\bnw \omega = 0
$,

\item homogeneous if
$
[\lvf,\bnw_X Y] - \bnw_{[\lvf,X]} Y - \bnw_X [\lvf Y] = 0
$
for all vector fields $X$, $Y$ on $\ctg Q$, where $\lvf$ denotes the Liouville vector field on $\ctg Q$.

\een

It turns out that torsionless, symplectic and homogeneous lifts are not unique, see e.g.\ \cite{BCGRS}. As shown in \cite{BNW}, one option to make the lift unique is to impose the additional condition that
$$
\omega\big(X_1,\Rbnw(Y,X_2)X_3 + \Rbnw(Y,X_3)X_2\big)
+
(\text{cyclic permutations of } X_1, X_2, X_3)
=
0
\,,
$$
where $\Rbnw$ denotes the curvature tensor of $\bnw$, viewed as a $2$-form on $\ctg Q$ with values in the $1,1$-tensor fields on $\ctg Q$. In \cite{BNW}, the authors used this connection to construct the Fedosov standard ordered star product on $\ctg Q$. Let us refer to this connection as the BNW lift of $\lcc$. Recently, the BNW lift was used in the context of homological reduction, see \cite{PRS}.

Starting from the classical papers of Yano and Patterson  \cite{YP1, YP2}, the problem of lifting geometric objects living on a manifold 
$Q$ to its cotangent bundle $\ctg Q$, or more generally to some tensor bundle over $Q$, has been addressed frequently, see 
\cite{Ganc, Kures, KM, Ocak} and further references therein. Very generally speaking, this problem is related to the discussion 
of natural geometric operations as treated in \cite{KMS}.

For our purposes, the notion of complete lift of a connection from $Q$ to $\ctg Q$ as invented in \cite{YP1} will be crucial. 
Let us denote this lift  by $\clcc$. It is defined as the Levi-Civit\'a connection of a certain pseudo-Riemannian metric $g^\lcc$ on $\ctg Q$, called the Riemann extension of $\lcc$, see below for the definition. 

After having completed this paper, we learnt that the result presented here was already obtained earlier by Pleba\'nski, Przanowski and Turrubiates \cite{PPT} using local coordinates. Nonetheless, it seems to us that our coordinate-free approach based upon the natural lifting of geometric objects in fibre bundles is worth communicating, especially from the point of view of applications in homological reduction of 
deformation quantized models with symmetries \cite{BHP,PRS}, see Section \ref{discussion} for a brief discussion of this application.


\section{Lifting operations for tensor fields}


Let us recall the following natural lifting operations turning objects on $Q$ into objects on $\ctg Q$ \cite{YP1,YP2}. Let $\langle \,\cdot\, , \,\cdot\, \rangle$ denote the natural pairing between vectors and covectors, and between $1$-forms and vector fields on $Q$. Every vector field $X$ on $Q$ defines a function $\tilde X$ on $\ctg Q$ by 
$$
\tilde X(p) := \langle p , X_{\pi(p)} \rangle
\,,\qquad
p \in \ctg Q
\,.
$$
We will refer to this function as the tautological function defined by $X$. The lift of a function $f$ on $Q$ to $\ctg Q$ is given by the pull-back under $\pi$,
$$
\mr v f := \pi^\ast f
\,.
$$
The lift of a $1$-form $\alpha$ on $Q$ to $\ctg Q$ is the vertical vector field $\mr v\alpha$ on $\ctg Q$ induced by the complete flow 
$$
\ctg Q \times \RR \to \ctg Q
\,,\qquad
(p,t) \mapsto p + t \alpha\big(\pi(p)\big)
\,.
$$
The lift of $1$-forms and the operation sending vector fields $X$ on $Q$ to their tautological functions $\tilde X$ on $\ctg Q$ combine to a lifting operation turning $1,1$-tensor fields $T$ on $Q$ into vector fields $\mr v T$ on $\ctg Q$. By definition, for $1,1$-tensor fields of the form $T = X \otimes \alpha$ with a vector field $X$ and a $1$-form $\alpha$,
$$
\mr v(X \otimes \alpha) = \tilde X (\mr v \alpha)
\,.
$$
There are two further lifting operations turning vector fields $X$ on $Q$ into vector fields on $\ctg Q$, the complete lift $\mr c X$ induced by the natural symplectic structure of $\ctg Q$ and the horizontal lift $\mr h X$ defined by the Levi-Civit\'a connection for any tensor bundle over $Q$. The complete lift $\mr c X$ is the Hamiltonian vector field generated by the tautological function $\tilde X$, i.e., the unique vector field on $\ctg Q$ satisfying
$$
\omega(\cdot , \mr c X)
=
\mr d \tilde X
\,.
$$
The horizontal lift $\mr h X$ is uniquely determined by the conditions 
\beq\label{G-hor}
\pi' \circ (\mr h X) = X \circ \pi
\,,\qquad
\Klcc \circ (\mr h X) = 0
\,,
\eeq
where $\Klcc : \tg(\ctg Q) \to \ctg Q$ is the connection mapping of $\lcc$, see eg.\ Section 1.5 in \cite{Buch2}. 

Let us derive the relation between $\mr c X$ and $\mr h X$. Computations are simplified by the following observation.

\ble\label{L-tautfn}

If two vector fields $V$ and $W$ on $\ctg Q$ coincide on all tautological functions defined by vector fields on $Q$, then $V=W$.

\ele

A proof using local coordinates was given in \cite{YP1}.

\bbw

It suffices to show that for all tangent vectors $V_p$ of $\ctg Q$ based at some $p$ outside the zero section, the following holds. If $V_p \tilde X = 0$ for all vector fields $X$ on $Q$, then $V_p = 0$. 

To prove this, let such $p$ and $V_p$ be given. Put $q = \pi(p)$ and choose a $1$-form $\alpha$ on $Q$ such that $\alpha(q) = p$. Then, $\alpha' \pi' V_p$ is based at $p$ and we may take the difference $V_p - \alpha' \pi' V_p$. Since $\pi'(V_p - \alpha' \pi' V_p) = 0$, there exists $\xi \in \ctg_q Q$ such that $V_p - \alpha' \pi' V_p$ is represented by the curve $t \mapsto p + t \xi$. It suffices to show that $\pi' V_p = 0$ and $\xi = 0$. For that purpose, we note that
\al{\label{G-L1-1}
(\alpha' \pi' V_p) \tilde X & = (\pi' V_p)\big(\alpha(X)\big)
\,,
\\ \label{G-L1-2}
(V_p - \alpha' \pi' V_p) \tilde X & = \xi(X_q)
}
for every vector field $X$ on $Q$. Let $X_q \in \tg_q Q$ be given. Using a chart and a bump function centered at $q$, we can extend $X_q$ to a vector field $X$ on $Q$ in such a way that $\alpha(X)$ is constant in some neighbourhood of $q$. Then, \eqref{G-L1-1} implies $(\alpha' \pi' V_p) \tilde X = 0$, so that \eqref{G-L1-2} yields $\xi(X_q) = 0$. Since this holds for all $X_q \in \tg_q Q$, we conclude that $\xi=0$. Now, \eqref{G-L1-1} and \eqref{G-L1-2} imply that 
\beq\label{G-L1-3}
(\pi' V_p)\big(\alpha(X)\big) = 0
\eeq
for all vector fields $X$ on $Q$. Let a smooth function $f$ on $Q$ be given. Using once again a chart and a bump function centered at $q$, we can construct a vector field $X$ on $Q$ such that $\alpha(X) = f$ in some neighbourhood of $q$. Then, \eqref{G-L1-3} yields that $(\pi' V_p) f = 0$. Since this holds true for all smooth functions on $Q$, we obtain $\pi' V_p = 0$. This yields the assertion.
\ebw

The following formulae will be needed throughout the paper.

\ble\label{L-lifts}

Let $X$, $Y$ be vector fields, $\alpha$ a $1$-form, and $T$ a $1,1$-tensor field on $Q$.

\ben

\sitem\label{i-L-lifts-pba}
$\tilde X \circ \alpha = \langle \alpha , X \rangle$.

\sitem\label{i-L-lifts-va}
$(\mr v \alpha) \tilde X = \mr v \langle \alpha , X \rangle$.

\sitem\label{i-L-lifts-vT}
$(\mr v T) \tilde X = \big(T(X)\big)^\sim$.

\sitem\label{i-L-lifts-pX}
$\pi' \circ (\mr c X) = X \circ \pi$.

\sitem\label{i-L-lifts-cX}
$(\mr c X) \tilde Y = \omega(\mr c X , \mr c Y) = [X,Y]^\sim$.

\een

\ele

\bbw

Points \rref{i-L-lifts-pba}--\rref{i-L-lifts-vT} are immediate. 

\rref{i-L-lifts-pX}.\abs 
We evaluate both sides at $p \in \ctg Q$ and apply them to a smooth function $f$ on $Q$. For the left hand side, this yields
$$
\pi'\big((\mr c X)_p\big) f
=
(\mr c X)_p (\mr v f)
\,.
$$
Let $H_{\tilde X}$ and $H_{\mr v f}$ denote the Hamiltonian vector fields of the functions $\tilde X$ and $\mr v f$, respectively. By \cite[Prop.\ 8.3.11]{Buch1}, we have $H_{\mr v f} = - \mr v(\mr d f)$. Using this and point \rref{i-L-lifts-va}, we find
$$
(\mr c X)(\mr v f)
=
H_{\tilde X} (\mr v f)
=
- H_{\mr v f} \tilde X
=
\mr v(\mr d f) \tilde X
=
\mr v (X f)
=
\pi^\ast(X f)
\,.
$$
Hence, 
$
\pi'\big((\mr c X)_p\big) f
=
X_{\pi(p)} f
$
for all $p$ and $f$. This yields the assertion.

\rref{i-L-lifts-cX}.\abs 
By definition of the complete lift,
\beq\label{G-L-lifts-cX-1}
(\mr c X) \tilde Y
=
\langle \mr d \tilde Y , \mr c X \rangle
=
\omega\big(\mr c X , \mr c Y)
\,.
\eeq
From this, we read off that
\beq\label{G-L-lifts-cX-2}
(\mr c X) \tilde Y
=
- (\mr c Y) \tilde X
\,.
\eeq
Then, we rewrite 
$$
\omega\big(\mr c X , \mr c Y)
=
\mr d \theta\big(\mr c X , \mr c Y)
=
(\mr c X) \langle \theta , \mr c Y \rangle
-
(\mr c Y) \langle \theta , \mr c X \rangle
-
\langle \theta , [\mr c X,\mr c Y] \rangle
\,.
$$
By point \rref{i-L-lifts-pX}, we have $\langle \theta , \mr c X \rangle = \tilde X$. According to Prop.\ 3.1.5 in \cite{Buch1}, point \rref{i-L-lifts-pX} also implies $\pi' \circ [\mr c X,\mr c Y] = [X,Y] \circ \pi$ and hence $\langle \theta , [\mr c X,\mr c Y] \rangle = [X,Y]^\sim$. In view of \eqref{G-L-lifts-cX-2}, this yields
$
\omega\big(\mr c X , \mr c Y)
=
2 (\mr c X) \tilde Y - [X,Y]^\sim
$.
The assertion now follows from \eqref{G-L-lifts-cX-1}.
\ebw

\bsz\label{S-ch}

For every vector field $X$ on $Q$, one has
$$
\mr h X = \mr c X + \mr v(\lcc X)
\,.
$$

\esz

\bbw

We have to show that the right hand side satifies the conditions \eqref{G-hor}, i.e.,
\beq\label{G-L-lifts-0}
\pi' \circ \big(\mr c X + \mr v(\lcc X)\big) = X \circ \pi
\,,\qquad
\Klcc\big(\mr c X + \mr v(\lcc X)\big) = 0
\,.
\eeq
The first condition follows from point \rref{i-L-lifts-pX} of Lemma \rref{L-lifts}. To prove the second condition, we use that for every vector field $X$ on $Q$, every $1$-form $\alpha$ on $Q$ and every $q \in Q$ one has \cite[Prop.\ 1.5.6]{Buch2}
\beq\label{G-L-lifts-1}
\Klcc(\alpha' X_q) = (\lcc_X \alpha)_q
\eeq
and that for every $p \in \ctg Q$, $\Klcc$ acts on the linear subspace $\tg_{p}(\ctg_{\pi(p)}Q) \subset \tg_{p}(\ctg Q)$ as the natural identification of that subspace with the fibre $\ctg_{\pi(p)} Q$. In view of the definition of the vertical lift of $1$-forms, the latter implies that
\beq\label{G-L-lifts-2}
\Klcc\big((\mr v \alpha)_p\big) = \alpha\big(\pi(p)\big)
\eeq
for all $1$-forms $\alpha$ on $Q$ and all $p \in \ctg Q$. To evaluate \eqref{G-L-lifts-1}, let $X$, $\alpha$ and $q$ be given and denote $p=\alpha(q)$. We claim that
\beq\label{G-L-lifts-3}
\alpha' X_q
=
(\mr c X)_p + \big(\mr v(\lcc X))_p + \big(\mr v(\lcc_X \alpha)\big)_p
\,.
\eeq
By Lemma \rref{L-tautfn}, it suffices to evaluate both sides on $\tilde Y$ for an arbitrary vector field $Y$ on $Q$. For the left hand side, point \rref{i-L-lifts-pba} of Lemma \rref{L-lifts} yields
$$
(\alpha' X_q) \tilde Y
=
X_q (\tilde Y \circ \alpha)
=
X_q \big(\alpha(Y)\big)
\,.
$$
For the right hand side, using in addition points \rref{i-L-lifts-va}, \rref{i-L-lifts-vT} and \rref{i-L-lifts-cX} of that lemma and the fact that $\lcc$ is torsion-free, we find
\ala{
\left\{
(\mr c X)_p + \big(\mr v(\lcc X)\big)_p + \big(\mr v(\lcc_X \alpha))_p
\right\}
\tilde Y
& =
[X,Y]^\sim(p) 
+ 
(\lcc_Y X)^\sim(p) 
+ 
\big(\mr v \langle \lcc_X \alpha , Y \rangle\big)(p)
\\
& =
(\lcc_X Y)^\sim(p) 
+ 
\big(\mr v \langle \lcc_X \alpha , Y \rangle\big)(p)
\\
& =
\big(\mr v \langle \alpha , \lcc_X Y \rangle\big)(p)
+ 
\big(\mr v \langle \lcc_X \alpha , Y \rangle\big)(p)
\\
& =
\big(\mr v (X \langle \alpha , Y \rangle)\big)(p)
\\
& =
X_q \langle \alpha , Y \rangle
\,.
}
This proves \eqref{G-L-lifts-3}. Next, \eqref{G-L-lifts-2} yields
\beq\label{G-L-lifts-4}
\Klcc\left(\big(\mr v (\lcc_X \alpha)\big)_p\right) = (\lcc_X \alpha)_q
\,.
\eeq
Now, applying $\Klcc$ to both sides of eq.\ \eqref{G-L-lifts-3} and using \eqref{G-L-lifts-1} and \eqref{G-L-lifts-4}, we obtain that the second condition in \eqref{G-L-lifts-0} holds true.
\ebw


\section{BNW lift and complete lift of the Levi-Civit\'a connection}


In the sequel, it will be convenient to view $1$-forms as mappings $\tg Q \to \RR$ and $1,1$-tensor fields as mappings $\tg Q \to \tg Q$. According to \cite{BNW}, the BNW lift of $\lcc$ is given by 
\al{\label{G-D-bnw-1}
\bnw_{\mr v \alpha} (\mr v \beta)
 & :=
0
\,,\qquad
\nabla_{\mr v \alpha} (\mr h X)
 :=
0
\,,\qquad
\bnw_{\mr h X}\left(\mr v \alpha\right)
 :=
\mr v\left(\lcc_X \alpha\right)
\,,
\\ \label{G-D-bnw-2}
\bnw_{\mr h X}(\mr h Y)
 & :=
\mr h \left(\lcc_X Y\right)
 +
\mr v
 \left(
\frac 1 2 \Rlcc(X,Y)
 + 
\frac 1 6 \Rlcc(X,\cdot)Y
 +
\frac 1 6 \Rlcc(Y,\cdot)X
 \right)
 }
for all vector fields $X$, $Y$ on $Q$ and $1$-forms $\alpha$, $\beta$ on $Q$, where $\Rlcc$ denotes the curvature tensor of $\lcc$. The complete lift of $\lcc$ will be denoted by $\clcc$. According to \cite{YP1}, this is the Levi-Civit\'a connection of the pseudo-Riemannian metric $g^\lcc$ on $\ctg Q$ given by 
\ala{
g^\lcc(\mr v \alpha,\mr v \beta) = 0
\,,\qquad
g^\lcc(\mr v \alpha,\mr c X) = \mr v\big(\alpha(X)\big)
\,,\qquad
g^\lcc(\mr c X,\mr c Y) = - \mr v\big(\lcc_X Y + \lcc_Y X\big)
}
for all vector fields $X$, $Y$ on $Q$ and $1$-forms $\alpha$, $\beta$ on $Q$. This metric is referred to as the Riemann extension of $\lcc$ in \cite{YP1}. Explicitly,
\ala{
\clcc_{\mr v \alpha} (\mr v \beta)
 & =
0
\,,\qquad
\clcc_{\mr v \alpha} (\mr c X)
 =
- \mr v(\alpha \circ \lcc X)
\,,\qquad
\clcc_{\mr c X} (\mr v \alpha)
 =
\mr v(\lcc_X \alpha)
\\
\clcc_{\mr c X} (\mr c Y)
 & =
\mr c(\lcc_X Y) 
+ 
\mr v\left(
\lcc X \circ \lcc Y + \lcc Y \circ \lcc X - \Rlcc(X,\cdot)Y - \Rlcc(Y,\cdot)X
\right)
}
for all vector fields $X$, $Y$ on $Q$ and $1$-forms $\alpha$, $\beta$ on $Q$. 

We will show that the BNW lift $\bnw$ arises from the complete lift $\clcc$ by symplectification in the sense of \cite{BCGRS}. We proceed by first rewriting $\clcc$ in terms of horizontal lifts and then applying the symplectification procedure. For that purpose, we need knowledge on how $\clcc$ acts on the vertical lifts of $1,1$-tensor fields on $Q$.

\ble\label{L-nablaT}

Let $X$ be a vector field on $Q$, $\alpha$ a $1$-form on $Q$ and let $T$, $S$ be $1,1$-tensor fields on $Q$.

\ben

\sitem\label{i-L-nablaT-aT}
$\clcc_{\mr v \alpha}(\mr v T) = \mr v(\alpha \circ T)$,

\sitem\label{i-L-nablaT-Ta}
$\clcc_{\mr v T}(\mr v \alpha) = 0$,

\sitem\label{i-L-nablaT-XT}
$\clcc_{\mr c X}(\mr v T) = \mr v(\lcc_X T) - \mr v(\lcc X \circ T)$,

\sitem\label{i-L-nablaT-TX}
$\clcc_{\mr v T}(\mr c X) = - \mr v (T \circ \lcc X)$,

\sitem\label{i-L-nablaT-TS}
$\clcc_{\mr v T}(\mr v S) = \mr v (T \circ S) $.

\een

\ele

\bbw

It suffices to prove all formulae for $T$ being of the form $T = Y \otimes \beta$ with a vector field $Y$ on $Q$ and a $1$-form $\beta$ on $Q$. In the computations, we use the properties of connection and the formulae of Lemma \rref{L-lifts}.

\rref{i-L-nablaT-aT}.\abs 
$
\clcc_{\mr v\alpha} \mr v(Y \otimes \beta)
=
\clcc_{\mr v\alpha} (\tilde Y \, \mr v \beta)
=
\big((\mr v \alpha) \tilde Y\big) \mr v \beta
=
(\mr v \langle \alpha , Y \rangle) \mr v \beta
=
\mr v (\alpha \circ Y \otimes \beta)
\,.
$

\rref{i-L-nablaT-Ta}.\abs 
$
\clcc_{\mr v (Y \otimes \beta)} (\mr v \alpha)
=
\clcc_{\tilde Y (\mr v \beta)} (\mr v \alpha)
=
\tilde Y \clcc_{\mr v \beta} (\mr v \alpha)
=
0
\,.
$

\rref{i-L-nablaT-XT}.\abs 
We find
$$
\clcc_{\mr c X} \big(\mr v(Y \otimes \beta)\big)
=
[X,Y]^\sim \, \mr v \beta + \tilde Y \mr v (\lcc_X \beta)
=
\mr v\left([X,Y] \otimes \beta + Y \otimes \lcc_X \beta\right)
\,.
$$
The first summand can be replaced by $\mr v(\lcc_X Y \otimes \beta - \lcc_Y X \otimes \beta)$. Thus
$$
\clcc_{\mr c X} \big(\mr v(Y \otimes \beta)\big)
=
\mr v\left(\lcc_X (Y \otimes \beta)\right) 
- 
\mr v\left(\lcc X \circ (Y \otimes \beta)\right)
\,.
$$

\rref{i-L-nablaT-TX}.\abs 
$
\clcc_{\mr v(Y \otimes \beta)}(\mr c X)
=
\tilde Y \clcc_{\mr v \beta}(\mr c X)
=
- \tilde Y \mr v (\beta \circ \lcc X )
=
- \mr v \big((Y \otimes \beta) \circ \lcc X\big)
$.

\rref{i-L-nablaT-TS}.\abs 
$
\clcc_{\mr v(Y \otimes \beta)}(\mr v S) 
=
\tilde Y \clcc_{\mr v \beta}(\mr v S) 
=
\tilde Y \, \mr v (\beta \circ S) 
=
\mr v \big((Y \otimes \beta) \circ S\big) 
$.
\ebw

Now, we are prepared for rewriting $\clcc$ in terms of horizontal lifts.

\ble\label{L-nch}

In terms of the horizontal lift operation, the complete lift of $\lcc$ is given by 
\ala{
\clcc_{\mr v \alpha} (\mr v \beta)
 & =
0
\,,\qquad
\clcc_{\mr v \alpha} (\mr h X)
=
0
\,,\qquad
\clcc_{\mr h X} (\mr v \alpha)
 =
\mr v(\lcc_X \alpha)
\\
\clcc_{\mr h X} (\mr h Y)
 & =
\mr h(\lcc_X Y) 
-
\mr v\left(\Rlcc(Y,\cdot)X\right)
}
for all vector fields $X$, $Y$ on $Q$ and $1$-forms $\alpha$, $\beta$ on $Q$. 

\ele

The last formula may be written more symmetrically in the form
\beq\label{G-nch-sym}
\clcc_{\mr h X} (\mr h Y)
 =
\mr h(\lcc_X Y) 
-
\frac 1 2 \mr v
\left(
\Rlcc(X,Y) + \Rlcc(X,\cdot)Y + \Rlcc(Y,\cdot)X
\right)
\,.
\eeq

\bbw

The first formula holds by definition of $\clcc$ and the second and the third formula follow immediately from the proposition and Lemma \rref{L-nablaT}. To prove the last formula, we compute 
$$
\clcc_{\mr h X} (\mr h Y)
 =
\mr h(\lcc_X Y) 
+
\mr v
\left(
\lcc Y \circ \lcc X 
- \lcc \lcc_X Y 
+ \lcc_X(\lcc Y) 
- \Rlcc(X,\cdot)Y 
- \Rlcc(Y,\cdot)X
\right)
\,,
$$
where the argument of $\mr v(\cdot)$ is a $1,1$-tensor field on $Q$. Evaluation of this term on $\tilde Z$ for some vector field $Z$ on $Q$ yields the tautological function of the vector field on $Q$ given by 
$$
\lcc_{\lcc_Z X} Y
- \lcc_Z \lcc_X Y 
+ \lcc_X \lcc_Z Y 
- \lcc_{\lcc_X Z} Y
- \Rlcc(X,Z)Y 
- \Rlcc(Y,Z)X
\,.
$$
Since $\lcc$ is torsion-free, the first $4$ terms combine to $\Rlcc(X,Z)Y$. This yields the last formula.
\ebw

Next, we symplectify $\clcc$ according to \cite{BCGRS}. For that purpose, we define a $1,2$-tensor field $N$ on $\ctg Q$ by
$$
\omega\big(N(V,W),U\big)
=
(\clcc_V \omega)(W,U)
$$
for all vector fields $U$, $V$, $W$ on $\ctg Q$. It is easy to check that 
\beq\label{G-D-Spf}
\slcc_V W := \clcc_V W + \frac 1 3 N(V,W) + \frac 1 3 N(W,V)
\eeq
defines a connection on $\ctg Q$ and that this connection is symplectic. To determine $\slcc$, we have to compute $N$. For that purpose, we have to evaluate $\omega$ on vertical lifts of $1$-forms and $1,1$-tensor fields on $Q$, and on horizontal lifts of vector field on $Q$.

\ble\label{L-ovh}

Let $X,Y$ be vector fields on $Q$, let $\alpha,\beta$ be $1$-forms on $Q$, and let $T,S$ be $1,1$-tensor fields on $Q$. 

\ben

\sitem\label{i-L-ovh-aa}
$
\omega(\mr v \alpha, \mr v \beta)
=
\omega(\mr v \alpha, \mr v T)
=
\omega(\mr v T, \mr v S)
=
0
$.

\sitem\label{i-L-ovh-aX}
$
\omega(\mr v \alpha, \mr h X)
=
\mr v \langle\alpha,X\rangle
$.

\sitem\label{i-L-ovh-XT}
$
\omega(\mr v T , \mr h X)
=
\big(T(X)\big)^\sim
$.

\sitem\label{i-L-ovh-XX}
$
\omega(\mr h X, \mr h Y)
=
0
$.

\een

\ele

\bbw

\rref{i-L-ovh-aa}.\abs This follows from the fact that the fibres of $\ctg Q$ are isotropic.

\rref{i-L-ovh-aX}.\abs We have
\beq\label{G-L-ovh-1}
\omega(\mr v \alpha , \mr h X)
=
(\mr v \alpha) \langle \theta , \mr h X \rangle
-
(\mr h X) \langle \theta , \mr v \alpha \rangle
-
\langle \theta , [\mr v \alpha,\mr h X] \rangle
\,.
\eeq
The second term vanishes, because $\mr v\alpha$ is vertical. Formula \eqref{G-hor} implies $\langle \theta , \mr h X \rangle = \tilde X$, so that point \rref{i-L-lifts-va} of Lemma \rref{L-lifts} yields $\mr v \langle \alpha , X \rangle$ for the first term. For the last term, we evaluate $[\mr v \alpha,\mr h X] \tilde Y$ for an arbitrary vector field $Y$ on $Q$. Decomposing $\mr h X$ according to the proposition and using the formulae of Lemma \rref{L-lifts}, we find 
$$
[\mr v \alpha,\mr h X] \tilde Y
=
\mr v 
\big(
\langle \alpha , [X,Y] \rangle
+
\langle \alpha , \lcc_Y X \rangle
-
X \langle \alpha , Y \rangle
\big)
\,.
$$
Since $\lcc$ is torsion-free, the terms on the right hand side combine to $- \mr v \langle \lcc_X \alpha , Y \rangle$. Thus,
$$
[\mr v \alpha,\mr h X]
=
- \mr v(\lcc_X \alpha)
\,.
$$
Since this is vertical, the last term in \eqref{G-L-ovh-1} vanishes, and the assertion follows. 

\rref{i-L-ovh-XT}.\abs It suffices to check this for $T = Y \otimes \alpha$ for any vectors field $Y$ on $Q$ and any $1$-form $\alpha$ on $Q$. By point \rref{i-L-ovh-aX}, 
$
\omega\big(\mr v (Y \otimes \alpha) , \mr h X\big)
=
\tilde Y \omega(\mr v \alpha , \mr h X)
=
\tilde Y \mr v \langle \alpha , X \rangle
$.
Using the formulae of Lemma \rref{L-lifts}, this can be rewritten as $\big((Y \otimes \alpha) (X)\big)^\sim$.

\rref{i-L-ovh-XX}.\abs Using the proposition and the fact that the fibres of $\ctg Q$ are isotropic, we can rewrite
$$
\omega(\mr h X , \mr h Y)
=
\omega(\mr c X , \mr c Y)
+
\omega\big(\mr v(\lcc X) , \mr h Y\big)
+
\omega\big(\mr h X , \mr v(\lcc Y)\big)
\,.
$$
By point \rref{i-L-ovh-XT} and the fact that $\lcc$ is torsion-free, the last two terms yield 
$$
(\lcc_Y X - \lcc_X Y)^\sim = [Y,X]^\sim
\,.
$$ 
By point \rref{i-L-lifts-cX} of Lemma \rref{L-lifts}, the first term evaluates to  $\omega(\mr c X , \mr c Y) = [X,Y]^\sim$. 
\ebw

\bbm

Point \rref{i-L-ovh-XX} states that the distribution on $\ctg Q$ consisting of the horizontal subspaces is isotropic (in fact, Lagrangian). It thus provides a Lagrangian complement to the Lagrangian distribution of the fibre tangent spaces. This comes as no surprise, as the Riemannian metric on $Q$ has a natural lift to $\ctg Q$ and the latter combines with the symplectic form to a K\"ahler structure on $\ctg Q$. 
\qeb

\ebm

Now, we can determine $N$.

\ble\label{L-N}

Let $X,Y$ by vector fields on $Q$ and let $\alpha,\beta$ be $1$-forms on $Q$. 

\ben

\sitem\label{i-L-N-aa}
$
N(\mr v \alpha,\mr v \beta)
=
N(\mr v \alpha,\mr h X)
=
N(\mr h X,\mr v \alpha)
=
0
$.

\sitem\label{i-L-N-XX}
$
N(\mr h X,\mr h Y)
=
2 \mr v\big(\Rlcc(Y,\cdot)X\big)
$.

\een

\ele

\bbw

For every combination of arguments, we have to compute $\omega\big(N(\cdot,\cdot),\mr v \gamma\big)$ for any $1$-form $\gamma$ on $Q$ and $\omega\big(N(\cdot,\cdot),\mr h Z\big)$ for every vector field $Z$ on $Q$. 

By definition of $N$ and the derivation property of connection,
\ala{
\omega\big(N(\mr v \alpha,\#_1),\#_2\big)
& =
\big(\clcc_{\mr v \alpha} \omega\big)(\#_1 , \#_2)
\\
& =
\mr v \alpha\big(\omega(\#_1 , \#_2)\big)
-
\omega\big(\clcc_{\mr v \alpha} \#_1 , \#_2 \big)
-
\omega\big(\#_1 , \clcc_{\mr v \alpha} \#_2 \big)
}
where $\#_1$ stands for $\mr v \beta$ and $\mr h X$ and $\#_2$ for $\mr v \gamma$ and $\mr h Z$. According to Lemmas \rref{L-nch} and \rref{L-ovh}, each of the terms on the right hand side vanishes, no matter what $\#_1$ and $\#_2$ are. Thus, $N(\mr v \alpha,\mr v \beta) = 0$ and $N(\mr v \alpha,\mr h X) = 0$. Analogous calculations yield
$
\omega\big(N(\mr h X,\mr v \alpha),\mr v \gamma\big)
=
0
$
and 
$$
\omega\big(N(\mr h X,\mr v \alpha),\mr h Z\big)
=
\mr v (X \langle \alpha , Z \rangle)
-
\mr v \langle \lcc_X \alpha , Z \rangle
-
\mr v \langle \alpha , \lcc_X Z \rangle
=
0
,
$$
due to the derivation property of connection. Here, we have also used that $(\mr h X)(\mr v f) = \mr v(X f)$ for all smooth functions on $Q$, which follows at once from the first of the defining relations for $\mr h X$ given in \eqref{G-hor}. Thus, $N(\mr h X,\mr v \alpha) = 0$. Finally, we find 
$$
\omega\big(N(\mr h X,\mr h Y),\mr v \gamma\big)
=
0
\,,\qquad
\omega\big(N(\mr h X,\mr h Y),\mr h Z\big)
=
2 \big(\Rlcc(Y,Z)X\big)^\sim
\,.
$$
Since 
$$
\omega\big(\mr v\big(\Rlcc(Y,\cdot)X\big),\mr v \gamma\big)
=
0
\,,\qquad
\omega\big(\mr v\big(\Rlcc(Y,\cdot)X\big),\mr h Z\big)
=
\big(\Rlcc(Y,Z)X\big)^\sim
\,,
$$
this yields the formula asserted for $N(\mr h X,\mr h Y)$.
\ebw

By plugging the formulae of Lemmas \rref{L-nch} and \rref{L-N}, together with \eqref{G-nch-sym}, into \eqref{G-D-Spf} and comparing the resulting formulae for $\slcc$ with \eqref{G-D-bnw-1} and \eqref{G-D-bnw-2}, we finally obtain

\btm

The BNW lift of $\lcc$ is obtained from the complete lift by symplectification in the sense of {\rm \cite{BCGRS}}.
\qed

\etm


\section{Discussion}
\label{discussion}


As already mentioned in the introduction, the BNW connection studied in this paper is useful in the theory of Fedosov deformation 
quantization. The main objective of this final section is to outline, why the BNW connection is particularly suitable  in homological reduction of deformation quantized models. 

Recall that the requirements of torsion freeness, symplecticity and homogeneity of the lifted connection, needed for the construction of the Fedosov star product, do not yet fix this connection uniquely.\footnote{We note that the Fedosov star products have been classified, both for the symplectic and for the Poisson case, see e.g. \cite{RW} and further references therein.} The BNW connection is one possible choice. Having chosen an admissible connection, e.g. the BNW connection, one constructs the Fedosov star product via a superderivation $D$ of antisymmetric degree, fulfilling $D^2 = 0$ and acting on the formal Weyl algebra bundle, see \cite{Fedosov} for the details. Often, $D$ is referred to as the Fedosov Abelian connection. For many purposes, it would be desirable to find an explicit formula for $D$ and, thus, also for the star product. This is, in general, a rather difficult task.  We refer to papers by Tosiek \cite{Tosiek1,Tosiek2}, where the reader can find sample calculations of the Abelian connection within some special geometries. In  \cite{BNW}, the Weyl and the standard ordered star products for the general cotangent bundle case are studied. The authors obtain results on the structure of the star product which boil down the problem of explicit calculation to the calculation of a certain bi-differential operator, see e.g.\ Proposition 6 therein for the standard order case. In \cite{PRS}, together with M. Pflaum,  we studied a model of gauge theory obtained via lattice approximation of Yang-Mills theory \cite{KR1,KR2, GR}. Here, the underlying classical phase space is the cotangent bundle over a Lie group endowed with a natural Lie group action. For this model, we calculated the BNW connection explicitly, see Propositions 3.3 and 3.4 therein. Moreover, building on results of \cite{BNW}, we were also able to calculate the Fedosov star product explicitly, see Proposition 3.10 in \cite{PRS}. 

In our model, the BNW lift has the pleasant and useful property of being $G$-invariant, see Proposition 3.5 in \cite{PRS}. We note in passing that this property holds true also for  the general cotangent bundle case (with the bundle  endowed with a lifted Lie group action). This follows rather immediately from the nature of the geometric construction provided in the present paper, e.g.\ by inspection of the equivariance properties of the horizontal and vertical lifting operators. The above invariance property turned out to be very useful in the study of homological reduction of our gauge model as presented in \cite{PRS}. We note that homological reduction  may be viewed as one possible quantum counterpart of singular symplectic reduction. As the reduced classical phase space is the space of classical physical observables, homological reduction yields a path towards studying quantum theory in terms of observables. So, let us briefly comment on this interesting approach. In \cite{FedosovLMP}, Fedosov had shown that there is a natural deformation quantization analog of classical regular symplectic reduction. Next, Bordemann, Herbig and Waldmann \cite{BHW} developed a deformation quantization formulation of the BRST-method. They proved that, under appropriate regularity properties of the group action, the BRST-procedure induces a star product on the reduced phase space. In \cite{BHP}, Bordemann, Herbig and Pflaum showed that this method may be extended to singular symplectic reduction of a  $G$-Hamiltonian system $(M, \omega, \Psi, J)$, provided  the following assumptions are fulfilled:

\begin{enumerate}

\item
The components of the moment map $J$ generate the vanishing ideal of the zero level set $J^{-1} (0)$.

\item
The Koszul complex on $J$ in the ring of smooth functions on phase space is acyclic. 

\item
 The star product of the underlying unreduced quantum deformation theory is $G$-invariant and $G$-covariant.

\end{enumerate}

Then one proceeds as follows, for details see \cite{BHP,PRS}.

\begin{enumerate}

\item
Given a $G$-Hamiltonian system $(M, \omega, \Psi, J)$, one constructs the classical BRST-complex $(\calA^\bullet,\calD)$ by taking the graded tensor product of the Chevalley--Eilenberg complex $\CE^\bullet (\g, \calC^\infty(M))$ associated with the $\g$-module $\calC^\infty (M)$ with the Koszul complex $(K^\bullet,\partial)$ on the moment map $J$ and endows it with the structure of  a differential graded commutative $\calC^\infty (M)$-algebra. Moreover, one shows that the latter carries a natural Poisson structure. Now, one can prove that, under the assumptions 1 and 2 above, the classical symplectically reduced space  is representable (via a deformation retract) as the zeroth cohomology of this BRST-complex  with its natural structure of a differential graded Poisson algebra.

\item 
Assume we are given a $G$-invariant and $G$-covariant star product $\star$ obtained by formal deformation quantization of the $G$-Hamiltonian system $(M, \omega, \Psi, J)$. Combining this star product with the natural product on the Gra{\ss}mann part, one can endow the $\CC[[\lambda]]$-module $\calA^\bullet[[\lambda]]$ of formal power series with values in $\calA^\bullet$ with a star product $\ast$. Moreover, one  constructs a deformation $\bm{\calD}$ of the classical BRST-differential, thus arriving at a formal deformation quantization $(\calA^\bullet[[\lambda]], \ast , \bm{\calD})$ (called the quantum BRST algebra) of the classical BRST algebra  $\calA^\bullet$. Finally, one can prove that there exists a deformed version of the contraction mentioned under point 1, giving rise 
to a star product on the symplectically reduced space. 

\end{enumerate}

In \cite{PRS}, we proved that the above conditions 1, 2 and 3 are fulfilled for our gauge model, indeed. For the proof of the two $G$-equivariance properties of the unreduced star product, the $G$-invariance of the BNW lift plays a basic role. 
Whereas the $G$-invariance property of the star product follows from quite general arguments, see Corollary 5.6 in \cite{PRS}, the $G$-covariance has to be rather checked by direct inspection, using the concrete form of the star product 
(see Proposition 5.9 in \cite{PRS}). So, here, using the BNW lift was very helpful. In particular, it was also important to know the BNW lift and the corresponding unreduced star product explicitly. 

It would be interesting to study whether there exist other torsionless, symplectic and homogeneous lifted connections, different from the BNW connection, which are $G$-invariant as well. This seems to be a difficult task, because there is no general theory of invariant connections over $G$-manifolds having a non-trivial orbit structure, see Section 1.9 in  
\cite{Buch2}.


\section*{Acknowledgements}


We are grateful to M.\ Dunajski for pointing out to us the paper of Pleba\'nski, Przanowski and Turrubiates \cite{PPT}. 

M.S.\ acknowledges funding by DFG under the grant SCHM 1652/2.

\end{document}